\documentclass[prc,twocolumn,superscriptaddress,twoside,floatfix]{revtex4-2}

\usepackage{amssymb,epsfig}
\usepackage{xcolor}

\usepackage{graphicx}
\usepackage[outdir=./]{epstopdf}

\begin{document}

\title{The $^{6}$H states studied in the $^2\text{He}(^8\text{He},^4\text{He})$ 
reaction and evidence of extremely correlated character of the $^{5}$H ground 
state}

\author{E.Yu.~Nikolskii}
\email{enikolskii@mail.ru}
\affiliation{National Research Centre ``Kurchatov Institute'', Kurchatov sq.\ 1,
123182 Moscow, Russia}
\affiliation{Flerov Laboratory of Nuclear Reactions, JINR,  141980 Dubna,
Russia}

\author{I.A.~Muzalevskii}
\affiliation{Flerov Laboratory of Nuclear Reactions, JINR,  141980 Dubna,
Russia}
\affiliation{Institute of Physics, Silesian University in Opava, 74601  Opava,
Czech Republic}

\author{A.A.~Bezbakh}
\affiliation{Flerov Laboratory of Nuclear Reactions, JINR,  141980 Dubna,
Russia}
\affiliation{Institute of Physics, Silesian University in Opava, 74601 Opava,
Czech Republic}

\author{V.~Chudoba}
\affiliation{Flerov Laboratory of Nuclear Reactions, JINR,  141980 Dubna,
Russia}
\affiliation{Institute of Physics, Silesian University in Opava, 74601  Opava,
Czech Republic}

\author{S.A.~Krupko}
\affiliation{Flerov Laboratory of Nuclear Reactions, JINR,  141980 Dubna,
Russia}

\author{S.G.~Belogurov}
\affiliation{Flerov Laboratory of Nuclear Reactions, JINR,  141980 Dubna,
Russia}
\affiliation{National Research Nuclear University ``MEPhI'', 115409 Moscow,
Russia}

\author{D.~Biare}
\affiliation{Flerov Laboratory of Nuclear Reactions, JINR,  141980 Dubna,
Russia}

\author{A.S.~Fomichev}
\affiliation{Flerov Laboratory of Nuclear Reactions, JINR,  141980 Dubna,
Russia}
\affiliation{Dubna State University, 141982 Dubna, Russia}

\author{E.M.~Gazeeva}
\affiliation{Flerov Laboratory of Nuclear Reactions, JINR,  141980 Dubna,
Russia}

\author{A.V.~Gorshkov}
\affiliation{Flerov Laboratory of Nuclear Reactions, JINR,  141980 Dubna,
Russia}

\author{L.V.~Grigorenko}
\affiliation{Flerov Laboratory of Nuclear Reactions, JINR,  141980 Dubna,
Russia}
\affiliation{National Research Nuclear University ``MEPhI'', 115409 Moscow,
Russia}
\affiliation{National Research Centre ``Kurchatov Institute'', Kurchatov sq.\ 1,
123182 Moscow, Russia}

\author{G.~Kaminski}
\affiliation{Flerov Laboratory of Nuclear Reactions, JINR,  141980 Dubna,
Russia}
\affiliation{Heavy Ion Laboratory, University of Warsaw, 02-093 Warsaw, Poland}

\author{M.~Khirk}
\affiliation{Skobeltsyn Institute of Nuclear Physics, Moscow State University,
119991 Moscow, Russia}
\affiliation{Flerov Laboratory of Nuclear Reactions, JINR,  141980 Dubna,
Russia}

\author{O.~Kiselev}
\affiliation{GSI Helmholtzzentrum f\"ur Schwerionenforschung GmbH, 64291
Darmstadt, Germany}

\author{D.A.~Kostyleva}
\affiliation{GSI Helmholtzzentrum f\"ur Schwerionenforschung GmbH, 64291
Darmstadt, Germany}
\affiliation{II. Physikalisches Institut, Justus-Liebig-Universit\"at, 35392
Giessen, Germany}

\author{M.Yu.~Kozlov}
\affiliation{Laboratory of Information Technologies, JINR,  141980 Dubna,
Russia}

\author{B. Mauyey}
\affiliation{Flerov Laboratory of Nuclear Reactions, JINR,  141980 Dubna,
Russia}
\affiliation{Institute of Nuclear Physics, 050032 Almaty, Kazakhstan}

\author{I.~Mukha}
\affiliation{GSI Helmholtzzentrum f\"ur Schwerionenforschung GmbH, 64291
Darmstadt, Germany}

\author{Yu.L.~Parfenova}
\affiliation{Flerov Laboratory of Nuclear Reactions, JINR,  141980 Dubna,
Russia}

\author{W.~Piatek}
\affiliation{Flerov Laboratory of Nuclear Reactions, JINR,  141980 Dubna,
Russia}
\affiliation{Heavy Ion Laboratory, University of Warsaw, 02-093 Warsaw, Poland}

\author{A.M.~Quynh}
\affiliation{Flerov Laboratory of Nuclear Reactions, JINR,  141980 Dubna,
Russia}
\affiliation{Nuclear Research Institute, 670000 Dalat, Vietnam}

\author{V.N.~Schetinin}
\affiliation{Laboratory of Information Technologies, JINR,  141980 Dubna,
Russia}

\author{A.~Serikov}
\affiliation{Flerov Laboratory of Nuclear Reactions, JINR,  141980 Dubna,
Russia}

\author{S.I.~Sidorchuk}
\affiliation{Flerov Laboratory of Nuclear Reactions, JINR,  141980 Dubna,
Russia}

\author{P.G.~Sharov}
\affiliation{Flerov Laboratory of Nuclear Reactions, JINR,  141980 Dubna,
Russia}
\affiliation{Institute of Physics, Silesian University in Opava, 74601 Opava,
Czech Republic}

\author{N.B.~Shulgina}
\affiliation{National Research Centre ``Kurchatov Institute'', Kurchatov sq.\ 1,
123182 Moscow, Russia}
\affiliation{Bogoliubov Laboratory of Theoretical Physics, JINR, 141980 Dubna,
Russia}

\author{R.S.~Slepnev}
\affiliation{Flerov Laboratory of Nuclear Reactions, JINR,  141980 Dubna,
Russia}

\author{S.V.~Stepantsov}
\affiliation{Flerov Laboratory of Nuclear Reactions, JINR,  141980 Dubna,
Russia}

\author{A.~Swiercz}
\affiliation{Flerov Laboratory of Nuclear Reactions, JINR,  141980 Dubna,
Russia}
\affiliation{AGH University of Science and Technology, Faculty of Physics and
Applied Computer Science, 30-059 Krakow, Poland}

\author{P.~Szymkiewicz}
\affiliation{Flerov Laboratory of Nuclear Reactions, JINR,  141980 Dubna,
Russia}
\affiliation{AGH University of Science and Technology, Faculty of Physics and
Applied Computer Science, 30-059 Krakow, Poland}

\author{G.M.~Ter-Akopian}
\affiliation{Flerov Laboratory of Nuclear Reactions, JINR,  141980 Dubna,
Russia}
\affiliation{Dubna State University, 141982 Dubna, Russia}

\author{R.~Wolski}
\affiliation{Flerov Laboratory of Nuclear Reactions, JINR,  141980 Dubna,
Russia}
\affiliation{Institute of Nuclear Physics PAN, Radzikowskiego 152, 31342
Krak\'{o}w, Poland}

\author{B.~Zalewski}
\affiliation{Flerov Laboratory of Nuclear Reactions, JINR,  141980 Dubna,
Russia}
\affiliation{Heavy Ion Laboratory, University of Warsaw, 02-093 Warsaw, Poland}

\author{M.V.~Zhukov}
\affiliation{Department of Physics, Chalmers University of Technology, S-41296
G\"oteborg, Sweden}

\date{\today.}

\begin{abstract}
The extremely neutron-rich system $^{6}$H was studied in the direct
$^2\text{H}(^8\text{He},{^4\text{He}})^{6}$H transfer reaction with a $26 A$ MeV
secondary $^{8}$He beam. The measured missing mass spectrum shows a broad bump
at $\sim 4-8$ MeV above the $^3$H+$3n$ decay threshold. This bump
can be interpreted as a broad resonant state in $^{6}$H at $6.8(5)$ MeV. The
population cross section of such a presumably $p$-wave state (or may be few
overlapping states) in the energy range from 4 to 8 MeV is
$d\sigma/d\Omega_{\text{c.m.}} \simeq 190^{+40}_{-80}$ $\mu$b/sr in the angular
range $5^{\circ}<\theta_{\text{c.m.}}<16^{\circ}$.
The obtained missing mass spectrum is practically free of the $^{6}$H events
below 3.5 MeV ($d\sigma/d\Omega_{\text{c.m.}} \lesssim 5$ $\mu$b/sr in the same
angular range). The steep rise of the $^{6}$H  missing mass spectrum at $\sim 3$
MeV allows to derive the lower limit for the possible resonant-state energy in
$^{6}$H to be $4.5(3)$ MeV. According to the paring energy estimates, such a
$4.5(3)$ MeV resonance is a realistic candidate for the $^{6}$H ground state
(g.s.). The obtained results confirm that the decay mechanism of the $^{7}$H
g.s.\ (located at 2.2 MeV above the $^{3}$H+$4n$ threshold) is the ``true'' (or
simultaneous) $4n$ emission. The resonance energy profiles and the momentum
distributions of fragments of the sequential $^{6}$H$ \,\rightarrow \,
^5$H(g.s.)+$n\, \rightarrow \, ^3$H+$3n$ decay were analyzed by the
theoretically-updated direct four-body-decay and sequential-emission mechanisms.
The measured momentum distributions of the $^{3}$H fragments in the $^{6}$H rest
frame indicate very strong ``dineutron-type'' correlations in
the $^{5}$H ground state decay.
\end{abstract}

\maketitle


\section{Introduction}


One of the important trends of the modern experimental nuclear physics, taking
advantages of the radioactive ion beam techniques, is the expansion of our
knowledge on nuclear systems located further beyond the proton and neutron
driplines. An important motivation here is the quest for the limits of
of nuclear structure existence: how far should we go beyond the driplines before
coming to situation when resonant structures become completely ``dissolved in
continuum''?

Recently, reliable spectroscopic information was obtained on the extreme
neutron-rich system $^{7}$H produced in the $^{2}$H($^{8}$He,$^{3}$He)$^{7}$H
reaction \cite{Bezbakh:2020,Muzalevskii:2021}. The $^{6}$H
population  in the $^{2}$H($^{8}$He,$^{4}$He)$^{6}$H reaction, which makes the
subject of the present work, is a natural byproduct of the above-mentioned
experiment.

Experimental information on the $^{6}$H resonant states is very limited. The
authors of Ref.\ \cite{Aleksandrov:1984} reported a value $E_T=2.7(4)$ MeV
(energy above the $^3$H+$3n$ decay threshold) for the $^6$H state produced in
the $^7$Li($^7$Li,$^8$B)$^{6}$H reaction. This result was confirmed
(with some reservations) in the $^9$Be($^{11}$B,$^{14}$O)$^{6}$H reaction
\cite{Belozyorov:1986}, giving $^{6}$H ground-state resonance energy
$E_T=2.6(5)$ MeV. The search for $^{6}$H in the $^{6}$Li($\pi^-$,$\pi^+$)
reaction was carried out in \cite{Parker:1990,Seth:1991}. No low-lying resonant
states were identified, which led the authors to conclusion that their results
cast serious doubt on the existence of the $^{6}$H resonance in the $0 - 5$ MeV
unbound region. The observation of the $^{6}$H resonant
states at $E_T=\{6.6(7)$, $10.7(7)$, $15.3(7)$, $21.3(4)\}$ MeV, populated in
the $^{9}$Be($\pi^-$,$pd)^6$H reaction, and $E_T=\{7.3(10)$, $14.5(10)$,
$22.0(10)$, $21.3(4)\}$ MeV states in the $^{11}$B($\pi^-$,$p^4$He$)^6$H
reaction was reported in Ref.\ \cite{Gurov:2007}. There was no indication of the
resonant state at 2.6-2.9 MeV in this work. The $^{6}$H g.s.\ energy
$E_T=2.9(9)$ MeV was determined in the $^8$He($^{12}$C,$^{14}$N)$^{6}$H reaction
\cite{Caamano:2008}. Our results are in contradiction with
\cite{Aleksandrov:1984,Belozyorov:1986,Caamano:2008} and are majorly in
agreement with \cite{Parker:1990,Gurov:2007}. We demonstrate in this work that
it is likely that the discussion about the actual position of the $^{6}$H g.s.\
is not finished yet, and it should be continued.

\begin{figure}
\begin{center}
\includegraphics[width=0.45\textwidth]{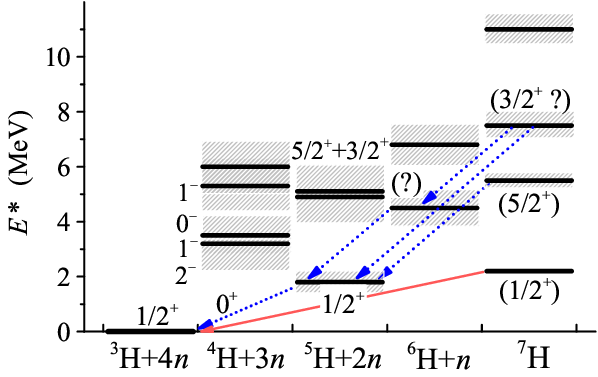}
\end{center}
\caption{The level schemes of $^{6}$H, and the known neighboring $^{4}$H,
$^{5}$H \cite{Korsheninnikov:2001,Golovkov:2004a,Golovkov:2005},  and $^{7}$H
\cite{Bezbakh:2020,Muzalevskii:2021}
systems, which are important for the discussions of this work. The solid red
arrow
illustrates the decay mechanism of $^{7}$H g.s.\ which is expected to be
``true'' $4n$ emission. The dotted blue arrows illustrate the decay mechanism of
the higher excitations in $^{7}$H, which is expected to be the sequential
$2n$+$2n$ and $2n$+$n$+$n$ emissions via the $^{5}$H and $^{6}$H excited states,
respectively. }
\label{fig:levels}
\end{figure}

The search for the $^{6}$H resonant states is an exciting challenge in itself,
however, here we face two important questions related also to our understanding
of neighboring systems.

\noindent (i) What are the decay mechanisms of the $^{7}$H ground ($E_T \approx
2.2$ MeV) and excited ($E_T \approx 5.5$ MeV) states? This is defined by the
spectra of its subsystems, see Fig.\ \ref{fig:levels}. For example, it could be
either the true $^{7}$H$\rightarrow ^{3}$H+$4n$ decay, or sequential
$^{7}$H$\rightarrow ^5$H(g.s.)+$2n$, or, else, the $^{7}$H$\rightarrow
^6$H(g.s.)+$n$ decay, depending on the ground state energies of $^{5}$H and
$^{6}$H. While for the $^{4}$H and $^{5}$H there are some relatively reliable
data, the spectrum of $^{6}$H is very uncertain.

\noindent (ii) What is the decay mechanism of the $^{6}$H ground state?
Intuitive vision of the situation, also confirmed by the theoretical estimates
of this work, tells us that the $^{6}$H g.s.\ decay is likely
to have a sequential $^{6}$H$\rightarrow \,^5$H(g.s.)+$n\rightarrow \,
^3$H+$3n$ character. Therefore, by studying the $^{6}$H decay, we may
also gain access to the decay properties of the $^{5}$H ground state. The
momentum distributions of the $^{3}$H fragment, measured in our experiment, can
be interpreted by assuming an unexpectedly strong ``dineutron'' correlation
character of the $^{5}$H ground state decay. The sequential $^{6}$H$\rightarrow
^5$H(g.s.)+$n\rightarrow ^3$H+$3n$ decay has never been studied before, and
interpretation of the data required extensive model studies and discussions of
this decay mechanism. Our results highlight the potential of the sequential
$^{6}$H$\rightarrow ^5$H(g.s.)+$n\rightarrow ^3$H+$3n$ decay as an important
source of information about the intermediate $^{5}$H system.

The data of this work obtained in the double $^{4}$He-$^{3}$H coincidences have
quite large statistics (among available data only for the
$^{9}$Be($\pi^-$,$pd)^6$H reaction \cite{Gurov:2007} statistics is better), and
it allows for a good MM energy resolution for the $^{6}$H spectrum and careful
treatment of the backgrounds. The consistent MM picture was obtained in the
triple $^{4}$He-$^{3}$H-$n$ coincidence data, which provide much smaller
statistical confidence, but which can be seen as practically background free.
Thus, the derived from our data detailed information on the low-energy spectrum
of $^{6}$H, shed light on the above-mentioned problems.


\section{Experiment}
\label{sec:exp}


The experiment was performed in the Flerov Laboratory of Nuclear Reactions
(JINR, Dubna) at the recently commissioned ACCULINNA-2 fragment-separator
coupled to the U-400M heavy ion cyclotron \cite{Fomichev:2018}. Recently the
$^{7}$H studies were carrid out in the $^{2}$H($^{8}$He,$^{3}$He)$^{7}$H
reaction \cite{Bezbakh:2020,Muzalevskii:2021}. The information on $^{6}$H is
naturally present in the data of these experiments due to a ``satellite''
$^{2}$H($^{8}$He,$^{4}$He)$^{6}$H reaction. The $^{7}$H experiments were
discussed in detail in Ref.\ \cite{Muzalevskii:2021}, and we only briefly sketch
here the information important for understanding of the $^{6}$H data.

\begin{figure}
\begin{center}
\includegraphics[width=0.48\textwidth]{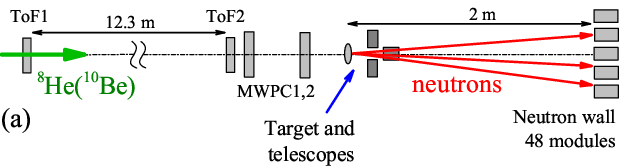}
\includegraphics[width=0.43\textwidth]{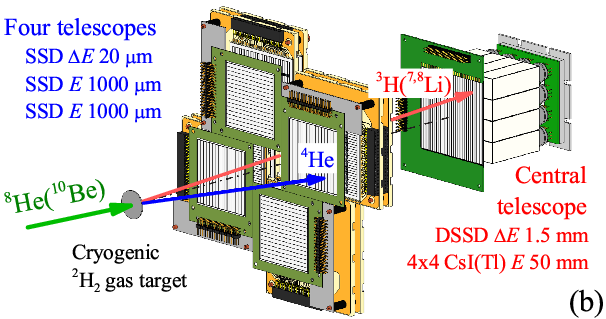}
\end{center}
\caption{Experimental setup for the $^{2}$H($^{8}$He,$^{4}$He)$^{6}$H and
$^{2}$H($^{10}$Be,$^{4}$He)$^{8}$Li reactions study: (a) general view and (b)
target and charged particle telescopes assembly.}
\label{fig:setup}
\end{figure}

The $^6$H system was produced in the $^2$H($^8$He,$^4$He)$^6$H reaction. The
secondary $^8$He beam was produced by the $33.4 A$ MeV $^{11}$B primary beam
fragmentation on 1 mm Be target. The $^8$He beam with intensity of $\approx 
10^5$ pps at $26 A$ MeV and $\sim 90 \%$ beam purity interacted with the 
deuterium nuclei in the cryogenic gas target, see Fig.\ \ref{fig:setup}. The 
target was 4 mm thick with 6 $\mu$m thick entrance and exit stainless steel 
windows. Being cooled down to 27 K, it had a thickness of $3.7 \times 10^{20}$
$\text{cm}^{-2}$.  The secondary beam diagnostics, made with the pair of thin
ToF plastics and the pair of position-sensitive chambers \cite{Kaminski:2020},
provided the determination of the hit position on the target and the time-of
flight measurement made for every individual beam ion with accuracy 1.8 mm and
280 ps, respectively.

The experimental setup, discussed in detail in Ref.\ \cite{Muzalevskii:2021},
involved four ``sideway'' $\Delta E$-$E$-$E$ telescopes detecting the recoil
nuclei ($^{4}$He in this experiment) emitted from the cryogenic deuterium gas
target in angular range $ 8^{\circ} - 26^{\circ}$ in laboratory system. The
20 $\mu$m thick, $50 \times 50$ mm$^2$ single-side Si front detector of the
telescope had 16 strips. Next to this $\Delta E$ detector was the 1 mm thick,
$61 \times 61$ mm$^2$ double-side Si strip detector having behind another 1 mm
thick veto detector. The ``central'' telescope, assigned for the registration of
the $^{3}$H fragments, originating from the $^6$H decay, consisted of the 1.5 mm
thick  $64 \times 64$ mm$^2$ double-side Si strip detector followed by the $4
\times 4$ array of CsI(Tl) scintillators. The charged particles, namely the
``fast'' decay tritons (with lab.\ energy $\sim 70\pm 30$ MeV) or Li isotopes
(in the case of the reference $^{2}$H($^{10}$Be,$^{4}$He)$^{8}\text{Li}^{\ast}$
reaction), were registered in the narrow forward cone $\theta \leq 6^{\circ}$,
with good angular ($\Delta \theta \leq 0.5^{\circ}$) and energy ($\Delta E/E
\leq 2 \%$) resolutions.

\begin{figure}
\begin{center}
\includegraphics[width=0.40\textwidth]{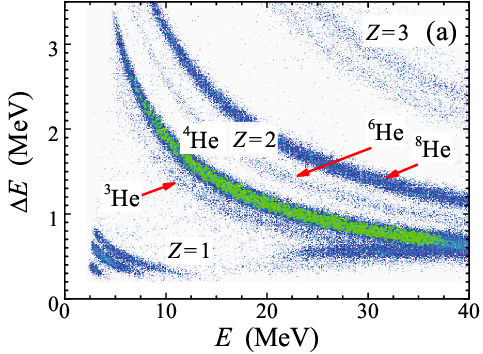}
\includegraphics[width=0.40\textwidth]{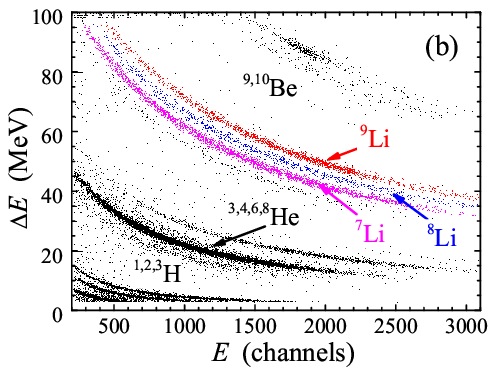}
\end{center}
\caption{Typical ID-plots. (a) For the sideway telescopes in the
$^{2}$H($^{8}$He,$^{4}$He)$^{6}$H reaction, where the green dots show the
$^{4}$He-$^{3}$H coincidences forming the $^{6}$H spectrum. (b) For the
central telescope in the $^{2}$H($^{10}$Be,$^{4}$He)$^{8}$Li reaction assigned
to calibration of the $^{6}$H MM spectrum.}
\label{fig:deltaee}
\end{figure}

The typical identification plots obtained with these detector telescopes are
illustrated in Fig.\ \ref{fig:deltaee}. A good quality of the helium isotopes'
identification is found, see panel (a), where the green dots show the
coincidences with ``fast'' tritons as well. The central telescope performance is
illustrated in Fig.\ \ref{fig:deltaee} (b) giving as example the
calibration $^{2}$H($^{10}$Be,$^{4}$He)$^{8}$Li reaction study. All hydrogen and
lithium isotopes are obviously well separated here.

\begin{figure}
\begin{center}
\includegraphics[width=0.42\textwidth]{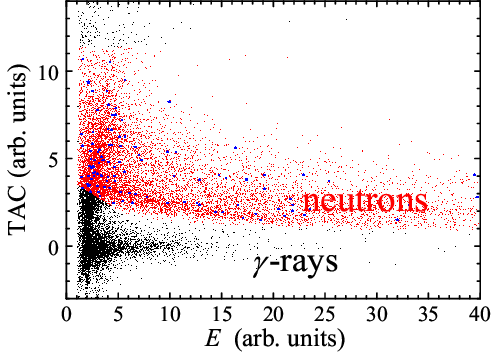}
\end{center}
\caption{The ID-plot for neutron spectrometer (time-to-amplitude ID parameter
vs.\ neutron energy). Red dots show events identified as neutrons and blue dots
indicate such events involved into the triple $^{4}$He-$^{3}$H-${n}$
coincidences.}
\label{fig:neutr-id}
\end{figure}

The setup also included a neutron wall consisting of 48 stilbene-crystal modules
where each 50-mm thick crystal was 80 mm in diameter \cite{Bezbakh:2018}. The
neutron wall was located near zero angle at a $\approx 2$ m distance from the
deuterium target, see Fig.\ \ref{fig:setup} (a). The neutron spectrometer
involved into the triple $^{4}$He-$^{3}$H-$n$ coincidences has a good
$n$-$\gamma$ separation provided by the so called pulse-shape analysis method
\cite{Bezbakh:2018}, see Fig.\ \ref{fig:neutr-id}.

The dedicated measurement with the $42 A$ MeV secondary $^{10}$Be beam (produced
from the $50 A$ MeV $^{15}$N primary beam) was performed to provide independent
calibration of the setup, see Section
\ref{sec:8li}.


\subsection{The Monte-Carlo simulations}
\label{sec:mc}


Complete Monte-Carlo (MC) simulations of the experimental setup for the
$^2\text{H}(^8\text{He},{^4\text{He}})^{6}$H reaction were performed. The
$^{6}$H MM resolution of the experiment $\sim 0.8-1.7$ MeV was determined in
different kinematical ranges, see Table \ref{tab:mc-resol} for details. The
maximum efficiencies (at $E_T \sim 5-7$ MeV) of the double $^4$He-$^3$H and
triple $^4$He-$^3$H-$n$ coincidence detection were $\sim 65 \%$ and $\sim 4 \%$,
respectively.
The setup efficiency as function of MM and the reaction center-of-mass angle,
$\theta_{\text{c.m.}}$, is demonstrated in Fig.\ \ref{fig:interpret} (a).
Important qualitative results of these studies are like follows.

\noindent (i) In the MM energy range of interest $E_T \sim 3-10$ MeV the setup
efficiency is both high and monotonous. The largest variation of efficiency is
quite modest, e.g.\ $\sim 40 \%$.

\noindent (ii) At about $\theta_{\text{c.m.}} \sim 16^{\circ}$ the setup
efficiency abruptly drops down. Therefore, the data interpretation for
$\theta_{\text{c.m.}} > 16^{\circ}$ becomes problematic due to the high
sensitivity to details of possible efficiency correction procedures.

The detailed MC simulations of the theoretically-motivated $^{6}$H MM spectra
(see Fig.\ \ref{fig:interpret}) and the isolated $^{6}$H resonance, expected
according to Refs.\ \cite{Aleksandrov:1984,Belozyorov:1986,Caamano:2008} at $E_T
\approx 2.6$ MeV (see Fig.\ \ref{fig:gs-mc}), have been performed, as well as MC
studies of the calibration $^{2}$H($^{10}$Be,$^{4}$He)$^{8}$Li reaction (see
Fig.\ \ref{fig:8li-mm}).

\begin{table}[b]
\caption{The $^{6}$H MM energy resolution (in MeV) of the setup for the
$^2\text{H}(^8\text{He},{^4\text{He}})^{6}$H reaction according to MC
simulations. The resolution is shown as function of the $^{6}$H MM energy
(columns, in MeV) and center-of-mass reaction angle (rows, in degrees). Missing
values correspond to a near zero efficiency of the setup.}
\begin{ruledtabular}
\begin{tabular}[c]{ccccc}
$\theta_{\text{c.m.}}$  & 5 & 10 & 15 & 20  \\
\hline
$10^{\circ}$   & 1.7 & 1.3 & 1.0 & 0.8 \\
$20^{\circ}$   & 1.7 & 1.5 & 1.3 & 1.0 \\
$30^{\circ}$   &     &     & 1.4 & 1.2 \\
\end{tabular}
\end{ruledtabular}
\label{tab:mc-resol}
\end{table}


\subsection{The neutron-unstable $^{8}$Li spectrum populated in the
$^{2}$H($^{10}$Be,$^{4}$He)$^{8}$Li reaction}
\label{sec:8li}

\begin{figure}
\begin{center}
\includegraphics[width=0.43\textwidth]{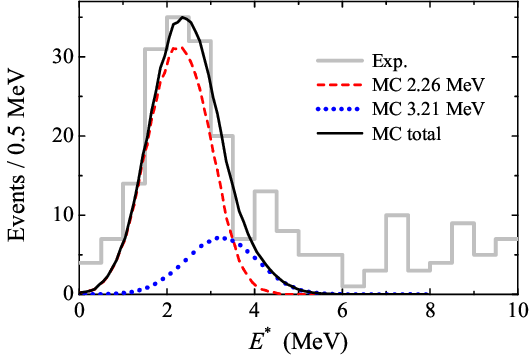}
\end{center}
\caption{The $^{8}$Li spectrum obtained in the
$^{2}$H($^{10}$Be,$^{4}$He)$^{8}\text{Li}^{\ast}$ reaction for $^4$He-$^7$Li
coincidence events.}
\label{fig:8li-mm}
\end{figure}

Dedicated test experiment was performed with the $^{10}$Be beam for the setup
calibration. The known neutron-unstable $^8$Li states were populated in the
$^2$H($^{10}$Be,$^4$He)$^8$Li reaction. The $\Delta E$-$E$ plots viewed in the
case where the $42 A$ MeV $^{10}$Be nuclei bombarded the deuterium gas target
were even more filled in their $^4$He loci than in the case of the $^8$He
projectiles. However, by imposing the $^4$He-Li coincidence condition this
background was strongly reduced. The first two particle-stable states of
$^{8}$Li (corresponding to the $^4$He-$^{8}$Li coincidence events) were found to
be poorly populated in this reaction. The first neutron-unstable resonant state,
known to be at the excitation energy of $E^{\ast}=2.26$ MeV \cite{www:nndc}, is
well seen in the $^4$He-$^{7}$Li coincidence events in Fig.\ \ref{fig:8li-mm}.
The energy of this state is measured with 250-keV error, but also, is well
interpreted by assuming that the next $^{8}$Li state with $E^{\ast}=3.21$ MeV is
populated with relative $\sim 30 \%$ probability, as predicted in the FRESCO
calculations.

\section{The $^{6}$H data}
\label{sec:data}


The $^{4}$He-$^{3}$H coincidence data (4650 events in total) obtained in the
$^2\text{H}(^8\text{He},{^4\text{He}})^{6}$H reaction are shown in Fig.\
\ref{fig:exp-all-6h}. The setup of experiment \cite{Muzalevskii:2021} was
optimized for the $^{7}$H search in the
$^2\text{H}(^8\text{He},{^3\text{He}})^{7}$H reaction, and, therefore, it was
not optimal for the $^{6}$H {studies. For that reason a relatively narrow
center-of-mass (c.m.) angular range was available for the $^4$He recoils
originating from the $^2$H($^8$He,$^4$He)$^6$H reaction, see Fig.\
\ref{fig:exp-all-6h} (b). Background conditions were quite poor for these
recoils because of random coincidences with alphas originating from other
intense reaction channels. This background can be seen in Fig.\
\ref{fig:exp-all-6h} (a) as the strong population of the $\{E_T,E_{3\text{H}}\}$
plane beyond the kinematical limit for the $^2$H($^8$He,$^4$He)$^6$H reaction
($E_{3\text{H}}$ is the $^3$H energy in the $^{6}$H c.m.\ frame).
The background in the low-energy part of the MM spectrum can be drastically
reduced by gating the data in the kinematically allowed range
$E_{\text{3H}}<E_T/2$ on the $\{E_T,E_{3\text{H}}\}$
plane. This selection results in 3850 events shown by red dots in Fig.\
\ref{fig:exp-all-6h} (b). The $^6$H MM spectrum derived from these events [blue
histogram in Fig.\ \ref{fig:exp-all-6h} (c)] shows a rise in the region
beginning at $E_T = 3.0 - 3.5$ MeV and going up to $E_T = 6$ MeV, where the
spectrum remains flat within the energy range extending up to $E_T = 9$ MeV. The
rate of this rise, coming to the flat top, matches well the shape characteristic
for relatively broad $p$-wave resonant states, as can be expected for
$^6$H. This rate is much faster than one may expect in situation without
resonant contributions [for example, the 4-body phase volume case is illustrated
by the orange dotted curve in Fig.\ \ref{fig:exp-all-6h} (c)]. This specific
shape of the MM spectrum allows us to claim that there is a resonance state, or
a group of overlapping resonance states in $^6$H located at MM energy $E_T 
\approx
6.8$ MeV.

\begin{figure}
\begin{center}
\includegraphics[width=0.48\textwidth]{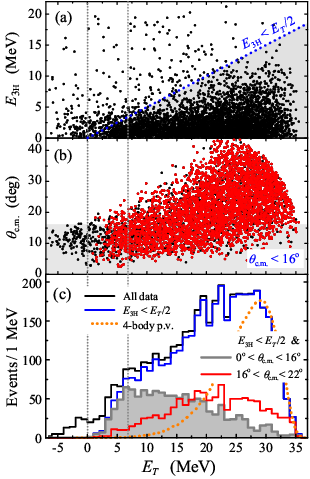}
\end{center}
\caption{Data on the $^{4}$He-$^{3}$H coincidence events considered for the
ascertainment of the $^6$H MM energy spectrum. (a) Correlation between the
$^3$H energy in the $^{6}$H c.m.\ frame $E_{3\text{H}}$ and the $^{6}$H MM
energy $E_T$. The gray triangle, bounded by the blue dotted line, shows the
kinematically allowed region. (b) Correlation between the center-of-mass
reaction angle and the $^{6}$H MM energy.
The gray rectangle shows the $\theta_{\text{c.m.}}<16^{\circ}$ cutoff region.
(c)
The $^{6}$H missing mass spectra: complete data (black histogram), kinematical
cutoff
$E_{3\text{H}} <  E_T/2$ (blue histogram), and additional cutoff
$\theta_{\text{c.m.}}<16^{\circ}$ (filled gray  histogram) and
$16^{\circ}<\theta_{\text{c.m.}}<22^{\circ}$ (red histogram). The orange dotted
curve illustrates the 4-body phase volume $\sim E_T^{7/2}$ convoluted with the
setup bias. The vertical gray dotted lines indicate the $^{3}$H+$3n$ threshold
and the position of the 6.8 MeV $^{6}$H resonant state.}
\label{fig:exp-all-6h}
\end{figure}

The 6.8 MeV bump can be made more visible by limiting the reaction c.m.\ angular
range as $\theta_{\text{c.m.}}<16^{\circ}$, see the gray histogram in Fig.\
\ref{fig:exp-all-6h} (c). All the MM spectra gated by some
$\theta_{\text{c.m.}}$ bands with $\theta_{\text{c.m.}}>16^{\circ}$ show no
resonating behavior, only monotonous growth up to $E_T \sim 20 $ MeV [for
example, see the red histogram in Fig.\ \ref{fig:exp-all-6h} (c)]. Partly this
is due to the setup efficiency in the $E_T \sim 6.8$ MeV energy range, which
rapidly degrades at $\theta_{\text{c.m.}} \gtrsim 16^{\circ}$ and comes to zero 
at $\theta_{\text{c.m.}} \sim 22^{\circ}$. In contrast, the energy range $E_T
\gtrsim 10-15$ MeV for  $\theta_{\text{c.m.}}>16^{\circ}$ is strongly boosted
due to the setup geometry. This effect is well illustrated in Figs.\
\ref{fig:exp-all-6h} (b) and \ref{fig:interpret} (a).


\subsection{The $^{6}$H c.m.\ angular distribution}


The cross section of the $^2$H($^8$He,$^4$He)$^6$H reaction populating the
expected low-lying resonant states of $^{6}$H was calculated using the FRESCO
code for $\Delta l = 1$ momentum transfer. The calculations are analogous to
those performed in \cite{Muzalevskii:2021} for the $^2$H($^8$He,$^3$He)$^7$H
reaction with the ``standard'' parameter set. The obtained center-of-mass cross
section is shown in Fig.\ \ref{fig:fresco}. The cross section features a broad
peak at about $\theta_{\text{c.m.}}\sim 8^{\circ}$, the rapid fall after
$\theta_{\text{c.m.}}>14-16^{\circ}$, and the diffraction minimum around
$\theta_{\text{c.m.}}\sim 24^{\circ}$.

In paper \cite{Muzalevskii:2021} the ``standard'' parameter set for FRESCO
calculations was modified to explain the experimentally observed missing
population in the angular range $10^{\circ} < \theta_{\text{c.m.}}< 14^{\circ}$,
which was assumed to correspond to the diffraction minimum of the
$^2$H($^8$He,$^3$He)$^7$H reaction. The ``standard'' parameter set predicted
this diffraction minimum at $\theta_{\text{c.m.}} \sim 18^{\circ}$. Strong
absorption or extreme peripheral character of the reaction were suggested in
\cite{Muzalevskii:2021} to explain the low-angle shift of the diffraction
minimum. One may expect that such a parameter modification is needed also for
the $^2$H($^8$He,$^4$He)$^6$H reaction calculations. However, both (i) the
situation observed in Fig.\ \ref{fig:fresco} with diffraction minimum at about
$\theta_{\text{c.m.}}\sim 24^{\circ}$ and (ii) the hypothetic situation of the
diffraction minimum shifted to smaller c.m.\ angles are qualitatively consistent
with the observed in Fig.\ \ref{fig:exp-all-6h} (c) absence of the 6.8 MeV bump
in the experimental MM spectrum for $\theta_{\text{c.m.}}>16^{\circ}$: the
angular range $16^{\circ} < \theta_{\text{c.m.}}< 20^{\circ}$ may correspond
either to diffraction minimum for the $\Delta l = 1$ cross section, or to the
right slope of its low-angle forward peak.

\begin{figure}
\begin{center}
\includegraphics[width=0.44\textwidth]{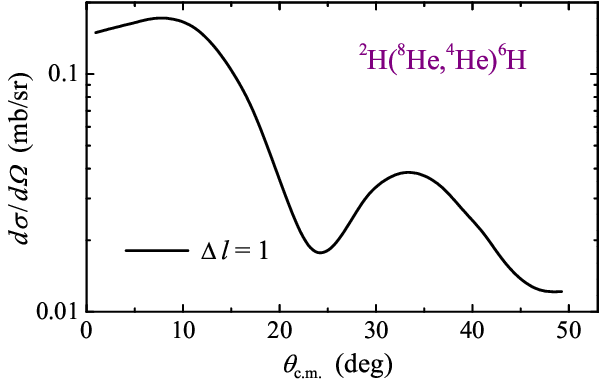}
\end{center}
\caption{The $\Delta l = 1$ cross section for the $^2$H($^8$He,$^4$He)$^6$H
reaction obtained in FRESCO calculations.}
\label{fig:fresco}
\end{figure}


\subsection{``Direct'' empty target subtraction}


\begin{figure}
\begin{center}
\includegraphics[width=0.48\textwidth]{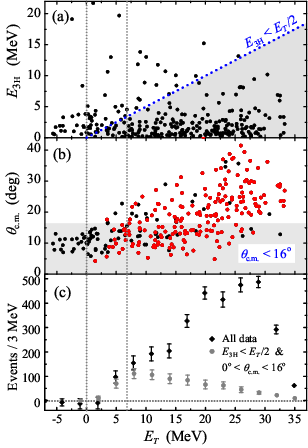}
\end{center}
\caption{Empty target data for the $^{4}$He-$^{3}$H coincidence events (see
Fig.\ \ref{fig:exp-all-6h} caption for more details) and background subtracted
$^{6}$H spectra (direct subtraction). (a) Correlation between the $^3$H energy
in the $^{6}$H c.m.\ frame $E_{3\text{H}}$ and the $^{6}$H MM
energy $E_T$. (b) Correlation between the center-of-mass
reaction angle and the $^{6}$H MM energy. (c) The $^{6}$H MM spectrum after
scaled empty target data subtraction: complete data (black diamonds),
kinematical cutoff $E_{3\text{H}} <  E_T/2$ and
$\theta_{\text{c.m.}}<16^{\circ}$ (gray circles). }
\label{fig:exp-empt-6h}
\end{figure}

The background contribution can be further reduced by taking into account the
empty target (deuterium gas out) data, see Fig.\ \ref{fig:exp-empt-6h}. The
empty target measurement collected around $17 \%$ of the beam integral providing
280 events in total and 190 events within the ``energy triangle'' of Fig.\
\ref{fig:exp-empt-6h} (a).

\begin{figure}
\begin{center}
\includegraphics[width=0.50\textwidth]{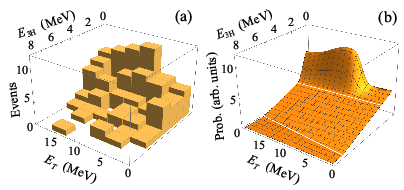}
\end{center}
\caption{(a) Empty target data in the correlation plane $\{E_T,E_{3\text{H}}\}$.
(b) Empty target data fit by a smooth analytical function.}
\label{fig:empty}
\end{figure}

Let's first use two simple procedures of the ``direct'' background subtraction
of the scaled empty target data. In the first case the total spectra are the
subject of subtraction, see Fig.\ \ref{fig:exp-empt-6h} (c), black diamonds.
Alternatively, the empty target spectrum is subtracted in the kinematical limits
$E_{3\text{H}} <  E_T/2$ and $\theta_{\text{c.m.}}<16^{\circ}$ (gray circles).
The two features should be pointed here:

\noindent (i) The subtraction spectra in the energy range $3.5 - 10$ MeV are
consistent with each other and consistent with the 6.8 MeV bump position as seen
in the spectrum without any background subtraction.

\noindent (ii) In both cases we get quite a low limit for the population of the
$E_T=0-3.5$ MeV energy range (the corresponding limits are $0 \pm 25$  events
and $10 \pm 9$ events). This point also is further discussed in Section
\ref{sec:does}.

So, the both direct subtraction methods produce consistent results and indicate
that we understand the nature of the apparatus-induced backgrounds in our
experiment.
However, because of the low statistics of the empty target data the  $E_T$ bin
size was to be set to quite a large value of 3 MeV. Despite the large bin size,
the statistical error bars produced by the two used procedures are quite large
and do not allow detailed quantitative conclusions. For that reason a more
stable background subtraction procedure is developed, which is based on the
smooth approximation of the empty target background data.


\subsection{``Advanced'' empty target subtraction}


\begin{figure*}
\begin{center}
\includegraphics[width=0.999\textwidth]{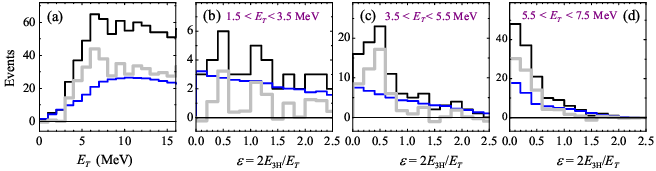}
\end{center}
\caption{Advanced empty target background subtraction. Initial $^{6}$H data
[black histogram, see gray histogram in Fig.\ \ref{fig:exp-all-6h} (c)], scaled
background (blue histogram), and corrected data (gray
histogram). Panel (a) shows the $^{6}$H MM spectrum. Panels (b), (c), and (d)
show the $\varepsilon$ distributions of the $^{3}$H fragment in the $^{6}$H rest
frame obtained in the MM  $E_T$ ranges $\{1.5,3.5\}$,  $\{3.5,5.5\}$, and
$\{5.5,7.5\}$ MeV, respectively.}
\label{fig:epsdis-all}
\end{figure*}

This subtraction procedure is based on assumption that the empty target
contribution is sufficiently smooth in the kinematical space.
The empty target data histogram in the  $\{E_T,E_{3\text{H}}\}$ plane is shown
in Fig.\ \ref{fig:empty} (a). This background has two components: the flat
component, weakly depending on energy, and the relatively narrow ``ridge'' at
small $E_{3\text{H}}$ values. It was approximated by a smooth analytical
function, see Fig.\ \ref{fig:empty} (b), and then a MC procedure was used to
subtract it from the data. The subtraction results obtained with the
empty-target data normalized to the $^8$He incoming beam flux are shown in Fig.\
\ref{fig:epsdis-all}. The motivation for the use of complicated
``two-dimensional'' background subtraction procedure and important conclusions
obtained as a result of this procedure are emphasized by the following two
issues.

\noindent (i) One may see in Figs.\ \ref{fig:epsdis-all} (c,d) that
the subtraction procedure reduces to zero the contributions in the kinematically
forbidden ranges $\varepsilon=2E_{3\text{H}}/E_T>1$ for the MM ranges
$\{3.5,5.5\}$ and $\{5.5,7.5\}$. This is a good indication that the background
subtraction procedure is reasonably well ``calibrated'' for the energy region of
interest.

\noindent (ii) The energy distribution in Fig.\ \ref{fig:epsdis-all} (b) is
flat, and there is no considerable event concentration in the kinematically
allowed range $\varepsilon<1$. If there is a flat background distribution in the
$\{E_T,E_{3\text{H}}\}$ plane for $E_T<3$ MeV, then, evidently, the
corresponding background contribution to the MM spectrum with the physical
kinematical selection $E_{3\text{H}}< E_T/2$ should be linear at $E_T<3$. This
is actually taking place, and, as a result, the whole $^{6}$H spectrum is
effectively reduced to zero in the MM range $E_T<3$ MeV, see Fig.\
\ref{fig:epsdis-all} (a).

The 6.8 MeV bump is clearly seen in the empty-target-corrected MM spectrum in
Fig.\ \ref{fig:epsdis-all} (a) with an average cross section of $\simeq 190(40)$
$\mu$b/sr being deduced for the c.m.\ angular range
$5^{\circ}<\theta_{\text{c.m.}}<16^{\circ}$.  This reasonably large cross
section is consistent with the resonant population mechanism. This value is also
in a very good agreement with the cross section obtained by FRESCO calculations,
see Fig.\ \ref{fig:fresco}. The steep rise of the spectrum at
$3.0-3.5$ MeV and the broad left tail of the 6.8 MeV bump provide together
an indication that another $^{6}$H state may be located at about 4.5 MeV, see
the discussion of Figs.\ \ref{fig:interpret} (c) and (d) in Section
\ref{sec:interpret} below. No indication on
the $2.6-2.9$ MeV state (the $^{6}$H ground state, as suggested in Refs.\
\cite{Aleksandrov:1984,Belozyorov:1986,Caamano:2008}) is found.


\subsection{Neutron coincidence data}


Practically background-free $^{6}$H data can be obtained by requesting
coincidence with one of the neutrons emitted in the $^{6}$H decay. The data on
the $^{4}$He-$^{3}$H-$n$ coincidence events (130 in total) from the
$^2\text{H}(^8\text{He},{^4\text{He}})^{6}$H reaction are shown in Fig.\
\ref{fig:exp-n-6h}. The background level of this spectrum can be estimated as
$\lesssim 3 \%$ from the ``kinematical triangles'' build for the $^{3}$H and
neutron emitted by $^{6}$H, see Figs.\ \ref{fig:exp-n-6h} (a) and (b). The c.m.\
angular distribution of the $^{4}$He-$^{3}$H-$n$ coincidence events is shown in
Fig.\ \ref{fig:exp-n-6h} (d). There is evidence that there is a peak at 6.8 MeV
in Fig.\ \ref{fig:exp-n-6h} (d), where indication on the 4.5 MeV structure can
be also found. There is no evidence of possible 2.6 MeV state in Fig.\
\ref{fig:exp-n-6h}: just one event is found in the 1.5 MeV energy bin around
$E_T =2.6$ MeV compared to the total 14 events within the $E_T \sim 3-8$ MeV MM
energy range, assigned to the broad 6.8 MeV peak.

\begin{figure}[tb]
\begin{center}
\includegraphics[width=0.46\textwidth]{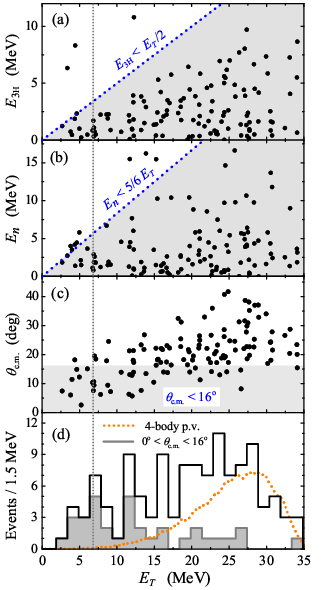}
\end{center}
\caption{The $^{4}$He-$^{3}$H-$n$ coincidence events. (a) Correlation between
the $^3$H energy in the $^{6}$H c.m.\ frame and
the $^{6}$H MM energy. (b) Correlation between the neutron energy in the
$^{6}$H c.m.\ frame and the $^{6}$H MM energy. The gray triangles in (a) and (b)
show the kinematically allowed regions. (c) Correlation between the
center-of-mass reaction angle and the $^{6}$H MM energy. The gray rectangle
shows the $\theta_{\text{c.m.}}<16^{\circ}$ cutoff region. (d) The $^{6}$H
missing mass spectrum gated by the kinematically allowed region of panel (a).
The vertical gray dotted line indicates the position of the 6.8 MeV $^{6}$H
resonant state. The orange dotted curve illustrates the 4-body phase volume
$\sim E_T^{7/2}$ convoluted with the setup bias.}
\label{fig:exp-n-6h}
\end{figure}

It is important to note that the neutron coincidence MM spectrum is nicely
described by the same curves as the empty-target-subtracted MM spectrum, see
Fig.\ \ref{fig:interpret}. This statement is, of course, valid within the much
larger statistical uncertainty of the neutron coincidence data.


\subsection{$^{6}$H spectrum interpretation}
\label{sec:interpret}

It should be carefully specified why and in which sense we speculate above about
the 6.8 MeV (and moreover, about the 4.5 MeV) states.

Possible interpretations of the low-energy $^{6}$H spectrum are illustrated
in Fig.\ \ref{fig:interpret}. In this figure the empty-target-corrected
$^{4}$He-$^{3}$H coincidence spectrum of Fig.\ \ref{fig:epsdis-all} (a) and the
$^{4}$He-$^{3}$H-$n$ coincidence spectrum of Fig.\ \ref{fig:exp-n-6h} (d) are
also corrected for the experimental efficiency by a MC procedure. For
consistency, the neutron coincidence spectrum in Fig.\ \ref{fig:interpret} (d) 
has the same $\theta_{\text{c.m.}}<16^{\circ}$ cutoff.

\begin{figure}[tb]
\begin{center}
\includegraphics[width=0.45\textwidth]{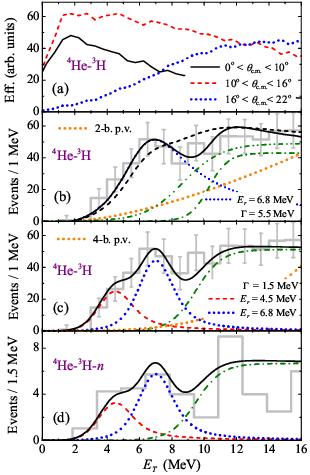}
\end{center}
\caption{(a) Efficiency of the $^{4}$He-$^{3}$H coincidence registration in
different $\theta_{\text{c.m.}}$ ranges. Possible interpretations of the $^{6}$H
spectrum with one broad (b) or two relatively narrow states (c) and (d). The
gray histograms in (b) and (c) show the efficiency-corrected $^{4}$He-$^{3}$H
coincidence data based on Fig.\ \ref{fig:epsdis-all} (a). The gray histogram in
(d) shows the  efficiency-corrected $^{4}$He-$^{3}$H-$n$ coincidence data based
on Fig.\ \ref{fig:exp-n-6h} (d). The red dashed and blue dotted curves
correspond to the possible contributions of the low-energy $^{6}$H states, the
green dash-dotted and dash-double-dotted curves are an option for the physical
background approximated by the Fermi-type profile. The black solid curves show
sum of all contributions. In panel (b) alternative fit without explicit deep at
$E_T\sim  9-10$ MeV is shown by the black dashed curve (the corresponding to it
physical background is shown by green dash-double-dotted curve). The 2-body
phase volume $\sim (E_T-E_{5\text{H}}^{(R)})^{3/2}$ for the $p$-wave
$^{5}$H(g.s.)+$n$ channel and the 4-body phase volume $\sim E_T^{7/2}$ and are
shown by the orange dotted lines in (b) and (c), respectively. }
\label{fig:interpret}
\end{figure}

The 4-body $^{3}$H+$n$+$n$+$n$ and 2-body $^{5}$H+$n$ phase volumes (orange
dotted curves) illustrate the possible profiles of nonresonant ``physical
backgrounds'' in Figures \ref{fig:interpret} (b) and (c).
We may see that such ``standard'' backgrounds have profiles which can not
explain the strong population of the $E_T \sim 3-8 $ MeV MM range. Some resonant
contributions are also needed.

The resonant cross section behavior at $E_T<9$ MeV is approximated by the
conventional Lorentz-like profiles
\[
\frac{d \sigma}{dE_T} \sim \frac{\Gamma(E_T)}{(E_r-E_T)^2+\Gamma(E_T)^2/4} \,,
\]
``corrected'' for the energy dependence of the width defined by Eq.\
(\ref{eq:dwid-seq}) below.

The interpretation with a single broad resonant peak is given in Fig.\
\ref{fig:interpret} (b), see the black solid curve. In these estimates we use
the width $\Gamma = 5.5$ MeV for the $E_T=6.8$ MeV resonant state, see Fig.\
\ref{fig:width}. This width value is likely to be the upper limit  for the
$^{6}$H resonant state, because the upper-limit parameters are used in the
estimates. For example, the maximum single-particle reduced width $\theta^2=1.5$
is used in Eq.\ (\ref{eq:wid-5h}) for the $^{5}$H-$n$ channel. For this
interpretation there is some indication for underestimation of the spectrum in
the low-energy region $E_T=3-5$ MeV. For a \textit{smaller width} of the
$E_T=6.8$ MeV resonant state or for a \textit{higher resonant energy} selection,
this underestimation becomes larger and is regarded as not acceptable.

Statistically, the deep in the experimental spectrum around $E_T=8-10$ MeV may
be regarded as not very significant. The ``smooth'' description of the data
without explicit resonant bump [see Fig.\ \ref{fig:interpret} (b), black dashed
curve] has $\chi^2$ value only somewhat larger than unity. This is much worse
than in the ``broad peak'' interpretation [black solid curve in Fig.\
\ref{fig:interpret} (b)], but statistically this is an acceptable value for the
$\chi^2$ criterion. However, for such a ``smooth'' fit we still need a resonant
state at $E_T=6.8$ MeV. For the ``smooth'' description of the data also the
resonance energy values $E_T<6.8$ MeV are acceptable. However, higher  resonance
energy values $E_T>6.8$ MeV are not acceptable due to the systematic
underestimation
of the low-energy data. Thus, the $E_T=6.8$ MeV resonance energy can be regarded
as an \textit{upper limit resonant energy} admissible for the data
interpretation with a \textit{single broad state}.

One may find in Fig.\ \ref{fig:interpret} (b), (c) that for the $^{6}$H MM
spectrum in the $E_T=4 - 8$ MeV energy range up to $\sim 35 \%$ of the 
population cross section can be related to ``physical background'' connected 
with low-energy tail
of the higher excitations. For that reason the lower limit given for the
uncertainty of the population cross section should be extended as
$d\sigma/d\Omega_{\text{c.m.}} \simeq 190^{+40}_{-80}$ $\mu$b/sr.


\subsection{Where is $^{6}$H ground state?}
\label{sec:gs}

Important feature of our data is the nonobservation of the $^{6}$H ground state
at $E_T=2.6-2.9$ MeV, as proposed in the earlier works
\cite{Aleksandrov:1984,Belozyorov:1986,Caamano:2008}. To quantify this fact we
performed complete MC simulations for the isolated ground state assuming the
$E_T=2.6$ MeV resonance energy and angular distribution predicted by
calculations of Fig.\ \ref{fig:fresco}. The MC simulations of our setup
efficiency, see Fig.\ \ref{fig:interpret} (a), show that this energy and angular
range $\theta_{\text{c.m.}}<16^{\circ}$ are the most favorable for such a
resonant state identification. Fig.\ \ref{fig:gs-mc} illustrates, which limits
on the population of the $E_T=2.6$ MeV resonance are imposed by our data. One
can see that even without any background subtraction this cross section limit
should be set as $d\sigma/d\Omega_{\text{c.m.}} \lesssim 25$ $\mu$b/sr. Such a
value is essentially smaller than $d\sigma/d\Omega_{\text{c.m.}} \sim 190$
$\mu$b/sr both observed around 6.8 MeV bump and predicted by calculations for
$\Delta l = 1$ angular momentum transfer. It is clear that whatever is the
applied background subtraction procedure, the actual cross section limit should
be smaller anyhow. With the direct empty target subtraction procedure the cross
section limit is obtained as $d\sigma/d\Omega_{\text{c.m.}} \lesssim 12$
$\mu$b/sr, see Fig.\ \ref{fig:exp-empt-6h} (c). According to the advanced
subtraction procedure the population is practically zero at $E_T<3.5$ MeV, see
Fig.\ \ref{fig:epsdis-all} (a). By assuming that the \emph{three} events,
appearing at $E_T<3.5$ MeV, could be attributed to such a state, the cross
section limit of its population is set as $d\sigma/d\Omega_{\text{c.m.}}
\lesssim 5$ $\mu$b/sr.

\begin{figure}[tb]
\begin{center}
\includegraphics[width=0.40\textwidth]{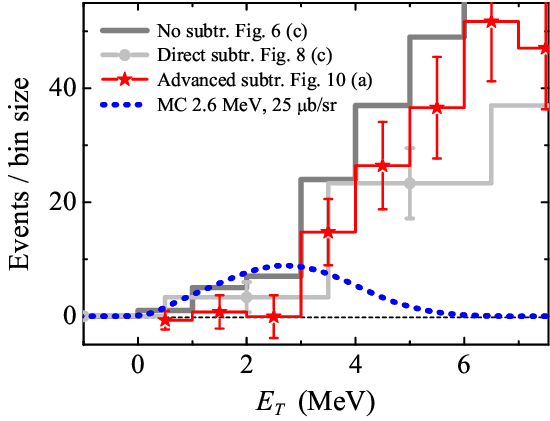}
\end{center}
\caption{MC simulations of the possible $^{6}$H g.s.\ resonance at $E_T=2.6$ MeV
(blue dotted curve). The gray histogram shows the data without any background
subtraction Fig.\ \ref{fig:exp-all-6h} (c). The red and light gray histograms
show the results of the direct and advanced empty target background subtraction,
see Fig.\ \ref{fig:exp-empt-6h} (c) and Fig.\ \ref{fig:epsdis-all} (a),
respectively.}
\label{fig:gs-mc}
\end{figure}

Here it is natural to ask the question: ``What is the lowest resonant energy
admissible by our data?''. We imply that the population rate for such a state
should be comparable for several possible low-lying states of $^{6}$H populated
by the $\Delta l = 1$ angular momentum transfer. It is discussed in the
theoretical estimates of Section \ref{sec:th} that much smaller widths of the
low-lying $^{6}$H states are
possible than it is assumed in Fig.\ \ref{fig:interpret} (b). For the $E_T=6.8$
MeV resonance with a smaller width (e.g., $\Gamma=1.5$ MeV) the interpretation
with two states, illustrated in Fig.\ \ref{fig:interpret} (c,d), is preferable.
The low-energy slope of the cross section can be described by a resonant state
with energy as low as $E_T=4.5$ MeV. This resonant contribution should be
interpreted as the \textit{lowest-energy resonant state} in $^{6}$H \textit{with
reasonably large population cross section}, which can be consistent with our
data.

Generally, one should keep in mind that more than two overlapping $^{6}$H states
may actually be expected in this energy range. Therefore, the ``two-state
situation'' in reality could mean ``more than one state''. The ground state
situation of $^{6}$H is further discussed in Section \ref{sec:dis}


\section{True and sequential decay of $^{6}$H}
\label{sec:th}


\begin{figure}[tb]
\begin{center}
\includegraphics[width=0.49\textwidth]{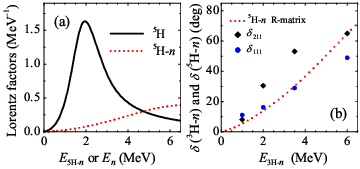}
\end{center}
\caption{(a) The Lorentz profiles, as used in Eq.\ (\ref{eq:dwid-seq}), for the
$^{5}$H(g.s.) subsystem and for the $^{5}$H-$n$ relative motion. (b) Phase
shifts, corresponding to the  $^{5}$H-$n$ profile in panel (a), are compared
with the experimental phase shifts of the lower spin-doublet states in $^{4}$H
with the $\{j,l,s\}$ sets  $\{1,1,1\}$ and $\{2,1,1\}$.}
\label{fig:prof-delta}
\end{figure}

The simplest idea about the character of the 4-body decay is based on
the phase volume (p.v.) approximation. The decay of such or analogous character
is often discussed as ``true 4-body decay'': there are no regions in the
momentum space which are emphasized by some forms of nuclear dynamics. The phase
space $dV_4$ of the 4-particle system can be defined by the three energies
$E_i=\varepsilon_i\,E_T$ corresponding to the three Jacobi vectors in momentum
space
\begin{equation}
dV_4 \sim E^{9/2}_T\, \delta(E_T-\textstyle \sum_i \varepsilon_i E_T)
\sqrt{\varepsilon_1 \varepsilon_2  \varepsilon_3} \,\,
d \varepsilon_1 d \varepsilon_2  d \varepsilon_3 \,.
\label{eq:pw4}
\end{equation}
The one-dimensional phase-volume energy distribution can be obtained by
integrating the phase space (\ref{eq:pw4}) over the two $\varepsilon$  variables
\begin{equation}
dV_{4}/d \varepsilon \sim E^{7/2}_T \,\sqrt{\varepsilon(1-\varepsilon)^4 } \,.
\label{eq:pw4i}
\end{equation}
This expression for the energy distribution is evidently the same for any of
the three Jacobi vectors. Therefore, it defines the single-particle energy
distributions both for $^{3}$H and $n$ fragments emitted in the $^{6}$H decay.

A more realistic scenario of the decay of $^{6}$H can be sequential process: the
emission of one neutron, which may lead to the population of the $^{5}$H ground
state. For theoretical modelling of the $^{6}$H sequential decay via the $^{5}$H
g.s.\ we employ the
generalization of the R-matrix-type expression, which was previously actively
used for the two-nucleon emission estimates in Refs.\
\cite{Grigorenko:2007a,Grigorenko:2009c,Olsen:2013,Brown:2015,Grigorenko:2015b}:
\begin{eqnarray}
\frac{\Gamma(E_T)}{d \varepsilon_{\text{5H}}} & = & \frac{E_T \langle V_3
\rangle ^2 }{2 \pi} \frac{\Gamma_{\text{5H-}n}(E_{\text{5H-}n})}
{(E_{\text{5H-}n}-E^{(R)}_{\text{5H-}n})^2 -
\Gamma^2_{\text{5H-}n}(E_{\text{5H-}n} )/4}  \qquad
\nonumber \\
& \times &
\frac{\Gamma_{5\text{H}}(E_{\text{5H}})}{(E_{\text{5H}}-E^{(R)}_{5\text{H}})^2 -
\Gamma^2_{5\text{H}}(E_{\text{5H}})/4} \,, \qquad
\label{eq:dwid-seq}\\
\langle V_3 \rangle ^2 & = & (E_T-E^{(R)}_{5\text{H}}-E^{(R)}_{n})^2  \qquad
\nonumber \\
& + &
[ \Gamma_{5\text{H}}(E^{(R)}_{\text{5H}})+
\Gamma_{5\text{H-}n}(E^{(R)}_{\text{5H-}n}) ]^2 /4\,,
\nonumber \\
E_{\text{5H}} & = &  \varepsilon_{\text{5H}} E_T \,,\quad    E_{\text{5H-}n} =
(1-\varepsilon_{\text{5H}}) E_T\,,
\nonumber
\end{eqnarray}
The $\Gamma_{5\text{H}}$ width dependence can be
parameterized as
\begin{equation}
\Gamma_{5\text{H}}(E_{5\text{H}}) = C_{5\text{H}} \, E_{5\text{H}}^2 \,, \qquad
C_{5\text{H}} = 0.5 \text{ MeV}^{-1}\,.
\label{eq:wid-5h}
\end{equation}
For $E^{(R)}_{5\text{H}}=1.8$ MeV this results in $\Gamma_{5\text{H}} = 1.62$
MeV, which is consistent with the data
\cite{Korsheninnikov:2001,Golovkov:2004a,Golovkov:2005}.
The neutron width can be defined by the standard R-matrix expression
\begin{equation}
\Gamma_{5\text{H-}n}(E_{5\text{H-}n}) = 2 \,  \frac{\theta^2}{2Mr^2_c} \,
P_{l=1}(E_{5\text{H-}n},r_c) \,,
\label{eq:wid-n}
\end{equation}
where $P_l$ is penetrability as a function of the decay energy $E_{5\text{H-}n}$
in the $^{5}$H+$n$ channel and its ``channel radius'' $r_c$. The Lorentz-type
profiles used in Eq.\ (\ref{eq:dwid-seq}) for the
$^{6}$H estimates are shown in Fig.\ \ref{fig:prof-delta} (a). They correspond
to the following parameters: $E_{5\text{H}}=2.25$ MeV, $E^{(R)}_{5\text{H-}n}=8$
MeV, $r_c= 3$ fm, and $\theta^2=1.5$. The phase shift in the $^{3}$H-$n$
channel, which can be associated with $\Gamma_{n}$ in Eq.\ (\ref{eq:wid-n}), is
shown in Fig.\ \ref{fig:prof-delta} (b): this can be seen as reasonably
consistent with phase shifts of the lowest states of $^{4}$H. The energy
distributions between the $^{5}$H(g.s.) and neutron, calculated by Eq.\
(\ref{eq:dwid-seq}), are illustrated in Fig.\ \ref{fig:epsdis-inp} (b) for two
$^{6}$H decay energies.

The $^{6}$H decay widths, estimated by Eq.\ (\ref{eq:dwid-seq}), is shown in
Fig.\ \ref{fig:width}, together with a trivial estimate of the $p$-wave neutron
emission on the $^{5}$H(g.s.) threshold made by Eq.\ (\ref{eq:wid-n}). For the
states with $E_T=4.5$ and 6.8 MeV the corresponding width values 3 and 5.5 MeV
are obtained. One may see that in proximity of the $^{5}$H(g.s.)-$n$ threshold
the width provided by the 4-body expression (\ref{eq:dwid-seq}) differs
qualitatively from that evaluated by  (\ref{eq:wid-n}). At higher energies the
difference becomes not so large. The 4-body expression provides result which is
somewhat smaller than the 2-body one (some part of the $^{5}$H continuum
strength described by a broad state remains outside the $^{6}$H decay energy
window).

\begin{figure}[tb]
\begin{center}
\includegraphics[width=0.49\textwidth]{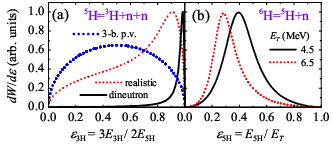}
\end{center}
\caption{The $\varepsilon$ energy distributions. (a) For the $^{5}$H(g.s.)
decay, between the $^{3}$H and $2n$ for different model assumptions about the
decay dynamics. (b) For for the $^{6}$H(g.s.) decay, between the $^{5}$H(g.s.)
and neutron.}
\label{fig:epsdis-inp}
\end{figure}

One should also note that the $^{6}$H g.s.\ may have quite low spectroscopic
factor of the $n$+$^5$H(g.s.) configuration. This idea comes
as analogy with the $^{7}$He g.s.\ situation, which can also be seen as a hole
in the neutron $p_{3/2}$ subshell from the shell model point of view. The
respective neutron spectroscopic factors of $0.3-0.6$ are typically derived or
predicted for the $^{7}$He g.s.\ (e.g.\ \cite{Renzi:2016,Fortune:2018}, and
Refs.\ therein). Therefore, the widths provided in Fig.\ \ref{fig:width} are
expected to be the \textit{upper limit estimates} for the widths, and one cannot
exclude that the actual widths of the $^{6}$H resonant states are much smaller.
The widths may be around $\Gamma \sim 1-3$ MeV, as assumed in Fig.\
\ref{fig:interpret} (c,d).

\begin{figure}[tbh]
\begin{center}
\includegraphics[width=0.44\textwidth]{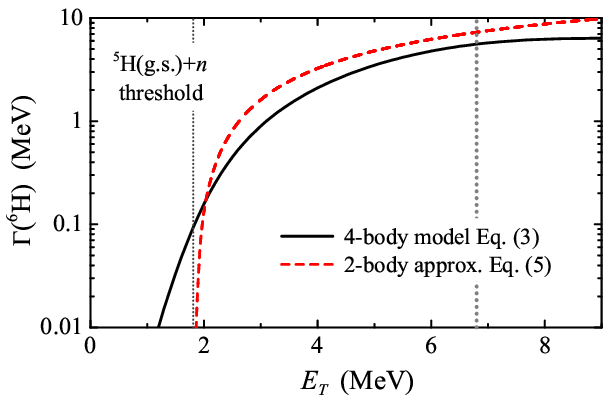}
\end{center}
\caption{The $^{6}$H g.s.\ width as function of the decay energy estimated by
trivial R-matrix expression (\ref{eq:wid-n}), assuming  2-body decay to the
$^{5}$H(g.s.) threshold, and by the 4-body sequential model of Eq.\
(\ref{eq:dwid-seq}).}
\label{fig:width}
\end{figure}


\subsection{Energy distributions of the decay products}


Though Eq.\ (\ref{eq:dwid-seq}) can be seen as a very simplistic model of the
$3n$ emission in $^{6}$H, it may provide some exclusive information, never
considered carefully before: the energy distributions of the decay products of
$^{6}$H may be calculated for more complicated dynamical assumptions than phase
volume.

For these calculations an additional input is required: the energy distribution
$^{3}$H+$n$+$n$ inside the $^{5}$H subsystem. Here we employ the
following three qualitatively different model distributions.

\noindent (i) ``3-b. p.v.''
--- three-body phase volume assumption about the decay of the $^{5}$H g.s. (the
standard uncorrelated assumption).

\noindent (ii) ``realistic'' --- the $^{5}$H g.s.\
energy distribution inspired by the experimental data \cite{Golovkov:2005}.

\noindent (iii) ``dineutron'' --- the highly correlated dineutron decay of the
$^{5}$H ground state.

These cases are illustrated in Fig.\ \ref{fig:epsdis-inp} (a). It
should be understood, what the above mentioned  ``inspired by experiment''
assumption means: the
energy distribution for $^{5}$H was reconstructed in \cite{Golovkov:2005} in the
energy range around the g.s.\ position (see Figs.\ 10 and 11 in
\cite{Golovkov:2005}). However, it is demonstrated in \cite{Golovkov:2005} that
the contribution of the broad $5/2^+-3/2^+$ doublet of excited $^{5}$H states
(located around $E_T \approx 5$ MeV) is large or even dominant in the $^{5}$H 
g.s.\
energy region ($E_T \approx 1.8$ MeV). For that reason
we can only guess or try to predict theoretically
\cite{Shulgina:2000,Grigorenko:2004,Grigorenko:2004a} what is the actual $^{5}$H
g.s.\ decay energy distribution.

\begin{table*}[bth]
\caption{Mean values of the $\varepsilon$ distributions of $^{3}$H fragments
obtained in the two $E_T$ decay energy ranges of $^{6}$H. The ``th.'' columns
show the theoretical results and the ``bias'' columns
give the corresponding values corrected for the experimental bias via the MC
procedure. The ``4-body p.v.'' is the four-body phase volume approximation of
the true $3n$ emission from $^{6}$H made by Eq.\ (\ref{eq:pw4i}). Models for the
sequential $^{6}$H$\, \rightarrow \, ^5$H(g.s.)+$n \, \rightarrow \, ^3$H+$3n$
decay: ``3-body p.v.'' --- the uncorrelated three-body phase volume decay of the
$^{5}$H g.s., ``realistic'' ---  the experiment-inspired distribution for the
$^{5}$H g.s., ``dineutron'' --- the highly correlated dineutron decay of the
$^{5}$H ground state. The column ``experiment'' shows the data from Figs.\
\ref{fig:exp-n-6h} and \ref{fig:epsdis-3h-final}. The experimental errors are
calculated by the MC procedure based on the available experimental statistics in
each case.}
\begin{ruledtabular}
\begin{tabular}[c]{ccccccccccc}
 Models: & \multicolumn{2}{c}{4-body p.v.} &  \multicolumn{2}{c}{3-body p.v.} &
 \multicolumn{2}{c}{realistic} & \multicolumn{2}{c}{dineutron} &
 \multicolumn{2}{c}{experiment} \\
ranges (MeV) & th. & bias & th. &  bias & th. & bias &  th. & bias &
$^{4}$He-$^{3}$H & $^{4}$He-$^{3}$H-$n$ \\
\hline
$3.5<E_T< 5.5$ & 1/3 & 0.30 & 0.29 & 0.28 & 0.33 & 0.30 & 0.43 & 0.39 & 0.42(3)
& 0.49(7) \\
$5.5<E_T< 7.5$ & 1/3 & 0.28 & 0.27 & 0.24 & 0.31 & 0.26 & 0.39 & 0.33 & 0.33(2)
& 0.24(8) \\
\end{tabular}
\end{ruledtabular}
\label{tab:3h-distr}
\end{table*}

By using the inputs from Figs.\ \ref{fig:prof-delta} and \ref{fig:epsdis-inp} we
obtain the energy distributions of the neutrons and $^{3}$H fragments in the
$^{6}$H rest frame, see Fig.\ \ref{fig:epsdis-n-th} and Fig.\
\ref{fig:epsdis-3h-final}, respectively. The estimated neutron distributions all
have a pronounced bimodal shape connected with the assumed sequential
$^{6}$H$\, \rightarrow \, ^5$H(g.s.)+$n \, \rightarrow\,  ^3$H+$3n$ mechanism
of the decay. Unfortunately, the single-neutron distribution is relatively
weakly sensitive to the decay mechanism of $^{5}$H, and the energy resolution of
the neutron spectrum in Fig.\ \ref{fig:exp-n-6h} is not sufficient to make
practical use of this information. In contrast, the $^{3}$H energy distribution
demonstrates strong sensitivity to the correlations in the $^{5}$H intermediate
system.

\begin{figure}[bth]
\begin{center}
\includegraphics[width=0.49\textwidth]{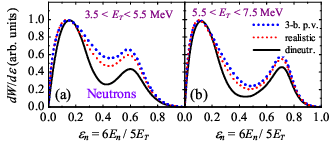}
\end{center}
\caption{The calculated $\varepsilon$ energy distributions of neutrons emitted
from the $^{6}$H states at $E_T = 4.5$ MeV (a) and  $E_T = 6.8$ MeV (b). The
black solid, red dashed, and blue dotted curves correspond to the ``dineutron'',
``realistic'', and 3-body phase volume models of the $^5$H g.s.\ decay,
respectively.}
\label{fig:epsdis-n-th}
\end{figure}

To make the above considerations quantitative, the $^{3}$H energy distributions
of Figs.\ \ref{fig:epsdis-3h-final} (a,b) were used in MC simulations, which
allowed us to take into account the bias of our experimental setup. The
resulting
distributions are shown in Figs.\ \ref{fig:epsdis-3h-final} (c,d), and the
numerical information about the $^{3}$H energy distributions is provided in
Table \ref{tab:3h-distr}.

\begin{figure}[tbh]
\begin{center}
\includegraphics[width=0.49\textwidth]{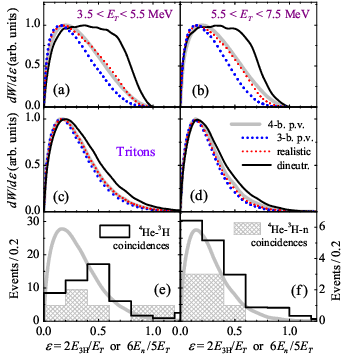}
\end{center}
\caption{The $\varepsilon$ energy distributions of $^{3}$H fragments emitted
from the  $^{6}$H states at $E_T = 4.5$ MeV (a,c,e) and  $E_T = 6.8$ MeV
(b,d,f).  The black solid, red dashed, and blue dotted curves correspond to the
``dineutron'',
``realistic'', and 3-body phase volume models of the $^5$H g.s.\ decay,
respectively. The thick gray curve shows the 4-body phase volume distribution
Eq.\ (\ref{eq:pw4i}). Panels (a,b) show the initial theoretical distributions,
while in panels (c,d) the experimental setup bias is taken into account via MC
procedure. Panels (e,f) show the experimental $\varepsilon$ energy distributions
for the $^{3}$H fragment in the $E_T$ energy ranges $\{3.5,5.5\}$ and
$\{5.5,7.5\}$ MeV. The black histograms show the distributions obtained from the
$^{4}$He-$^{3}$H coincidence data (left axis). The gray hatched histograms show
the distributions obtained from the $^{4}$He-$^{3}$H-$n$ coincidence data (right
axis). The MC 4-body phase volume distributions are shown by the thick gray
curves.}
\label{fig:epsdis-3h-final}
\end{figure}

The experimental energy distributions for the $^{3}$H fragment in the $^{6}$H
c.m.\ system for the $^{4}$He-$^{3}$H and $^{4}$He-$^{3}$H-$n$ coincidence
events are shown in Fig.\ \ref{fig:epsdis-3h-final} (e,f). These
distributions are consistent with each other within the available statistics in
the energy ranges $E_T=3.5-5.5$ MeV and $E_T=5.5-7.5$ MeV. One may conclude from
Fig.\ \ref{fig:epsdis-3h-final} and Table \ref{tab:3h-distr}
that the preferable interpretation of the data suggests the extremely correlated
emission of two neutrons from the $^{5}$H(g.s.) intermediate system.


\section{Discussion}
\label{sec:dis}


In this Section we are going further argument that the 6.8 MeV bump and cross
section rise at 4.5 MeV are likely to represent the actual ground and first
excited state (states), while the 2.6 MeV ground state, which is broadly
accepted now in the literature, has actually a very limited experimental
support.


\subsection{Does the 2.6 MeV state exist in the $^{6}$H?}
\label{sec:does}

The vision that the 2.6 MeV ground state energy of $^{6}$H is solidly
established is quite widespread. We have to point here that this misleading
impression is partly supported by some problem of data \emph{representation}
in the NNDC database. Namely, in NNDC page for the $^{6}$H level scheme one gets
information that the ground state population was taking place in \emph{all the
five} available reactions --- three in papers
\cite{Aleksandrov:1984,Belozyorov:1986,Caamano:2008} and two in
\cite{Gurov:2007}. Such a broad experimental support may look impressive.
However, if one opens the corresponding evaluation pdf files at NNDC
\cite{www:nndc} for the $^{6}$H population in pion absorption reactions ``B''
and ``D'' (based on Ref.\ \cite{Gurov:2007}), then it is easy to find
out that the presumed $^{6}$H g.s.\ at 2.6 MeV is actually not populated in
these reactions. The lowest-energy state, as obtained in this work, has energies
6.6(7) MeV for the $^{9}$Be($\pi^-$,$pd)^6$H reaction and 7.3(10) MeV for the
$^{11}$B($\pi^-$,$p^4$He$)^6$H reaction. These values are evidently in a nice
agreement with our 6.8(5) MeV bump.

The spectra shown in Ref.\ \cite{Gurov:2007} have good statistics (thousands of
events), comparable to statistics obtained in our experiment. The shapes of the
MM spectra are qualitatively the same as our MM spectrum: the energy region
under $E_T \sim 3$ MeV is poorly populated; then, there is a kink or bump in the
spectra at about $E_T \sim 7$ MeV; at higher energy the spectrum is reasonably
flat. Actually, two reasonably consistent sets of excited states are
additionally claimed in Ref.\ \cite{Gurov:2007}: $E_T=\{ 10.7(7)$, $15.3(7)$,
$21.3(4)\}$ MeV, populated in the $^{9}$Be($\pi^-$,$pd)^6$H reaction, and
$E_T=\{14.5(10)$, $21.3(4)\}$ MeV states populated in the
$^{11}$B($\pi^-$,$p^4$He$)^6$H reaction. We observe some oscillation in this
energy range in our spectrum with statistical significance analogous to that of
Ref.\ \cite{Gurov:2007}; however, we consider this  statistical significance as
insufficient to claim additional states in the spectrum of $^{6}$H.

The studies of the $^{6}$Li($\pi^-$,$\pi^+$)$^{6}$H reaction provided no
evidence for low-lying resonant states of $^{6}$H \cite{Parker:1990,Seth:1991}.
The authors performed the dedicated search for the 2.6 MeV ground state and
found that ``... In the missing mass region $0-5$ MeV $95\%$ confidence upper
limits of $0-5$ nb/sr for the production cross section were set.'' This limit
should be compared to the typical expected $^{6}$H g.s.\ population cross
section of $\sim 40$ nb/sr for this reaction. If we look directly at the missing
mass spectrum of the $^{6}$Li($\pi^-$,$\pi^+)^6$H reaction in
\cite{Parker:1990}, then some evidence for a kink in the spectrum can be seen at
$E_T \sim 7-9$ MeV, where the typical resonant population cross section of $\sim
30-70$ nb/sr is achieved. So, this data can be seen as being in qualitative
agreement with our result: there is no expressed bump in the cross section, but 
the typical resonant cross section is achieved at energies consistent with our
$E_T \approx 6.8$ MeV value.

Let's now have some critical review of the experiments in which the 2.6 MeV
g.s.\ was observed.

The $^{6}$H resonant state was reported for the first time in Ref.\
\cite{Aleksandrov:1984}. It gives the g.s.\ energy $E_T=2.7(4)$ MeV
for the $^6$H state produced in the $^7$Li($^7$Li,$^8$B)$^{6}$H reaction.
Actually a broad structure with $E_T=1.8-4.5$ MeV is observed, which
statistically is quite convincing ($\sim 300$ events). The data is strongly
contaminated with various backgrounds ($40-60 \%$ in the region of resonance
bump, according to \cite{Aleksandrov:1984}). Mechanism of this reaction is a
complicated ``bidirectional'' transfer $(-2p,+1n)$. Now we can point that
$^7$Li($^7$Li,$^9\text{B}^{*})^{5}$H(g.s.) can be  responsible for the formation
of this bump, where the $^{9}\text{B}^{*}$ excited states located somewhat above
the $^8$B+$n$ threshold are populated (for example, the $E^*=18.6$ or 20.7 MeV 
states). This is the much ``easier'' reaction (just $-2p$ transfer) and 
something like an order of the magnitude higher population cross section may be 
expected for it. The authors of \cite{Aleksandrov:1984} avoided this 
interpretation, as ``... the $^{5}$H nucleus is known not to exist''. Now the 
low-lying resonant g.s.\ of $^{5}$H with $E_T \approx 1.8$ MeV is solidly 
established and its population seem to be very favourable scenario for this 
reaction.

The $E_T=2.6(5)$ MeV bump was claimed to be observed in the
$^9$Be($^{11}$B,$^{14}$O)$^{6}$H reaction in Ref.\ \cite{Belozyorov:1986}.
There are problematic issues concerning this experiment. (i) Marginal statistics
and large backgrounds ($\sim 20$ events are spread on the top of $\sim 40$ of
expected background events). (ii) The obtained events are actually spread in a
much narrower energy range $E_T=2.1-3.1$ MeV than in \cite{Aleksandrov:1984}.
Point (ii) is probably partly connected with fact that the kinematical cut-off
for the $^9$Be($^{11}$B,$^{14}$O)$^{6}$H reaction in \cite{Belozyorov:1986} is
taking place at $E_T \approx  3.2- 3.5$ MeV leading to unknown strong 
distortions
of the MM spectrum in the region of expected $^{6}$H resonance peak. (iii) The
ground state of the neighboring $^{5}$H nuclide was not observed in Ref.\
\cite{Belozyorov:1986} in the analogous $^9$Be($^{11}$B,$^{15}$O)$^{5}$H
reaction (otherwise providing on average a reliable 10-fold larger statistics
than $^{6}$H data).

The $^{6}$H g.s.\ energy of $E_T=2.9(9)$ MeV was claimed in  \cite{Caamano:2008}
basing on the events which could originate from the
$^8$He($^{12}$C,$^{14}$N)$^{6}$H reaction. The problem is that there was no
channel identification in \cite{Caamano:2008}, which can reliably distinguish
among the $^{12}$N, $^{13}$N, and $^{14}$N recoils, and, consequently, among the
$^{7}$H, $^{6}$H, and $^{5}$H products. Assignment for each of these products
was solely based on assumption that only the low-lying near-threshold ground
state is populated in each case. Specifically for $^{6}$H in \cite{Caamano:2008}
there are \emph{five} events spread from 0 to 7.5 MeV excitation in the MM
spectrum and only \emph{three} events located between 1.5 to 5.5 MeV were
assigned as belonging to $^{6}$H g.s.\ resonance. This kind of data, taken
without interpretation, actually can be considered as not contradicting to our
data.

So, we can see that the $E_T \approx 2.6$ MeV state has very questionable 
support
in experimental data. All the experiments
\cite{Aleksandrov:1984,Belozyorov:1986,Caamano:2008} in which it was observed
have important experimental problems (statistical significance, channel
identification, etc.). In contrast, in experiments with large statistics, clear
channel identification and background treatment (Refs.\
\cite{Parker:1990,Gurov:2007} and our data) one gets the first expressed
lowest-energy feature at $E_T \approx 7$ MeV.
It is, of course, not impossible, that the $\approx 2.6$ MeV state was 
specifically
poorly populated in the reactions of Ref.\ \cite{Parker:1990,Gurov:2007} and our
reaction. However, such suggestion is quite unnatural, as there should be
several excited states of $^{6}$H (all populated by the same $\Delta l=1$ as the
ground state) within several MeV of excitation. All of them are expected to be
populated comparably with the ground state, while here we find the 6.8 MeV bump,
having typical for direct single-step transfer reactions cross sections (e.g.\
$d\sigma/d\Omega_{\text{c.m.}} \simeq 190$ $\mu$b/sr in our reaction).

We suggest to consider the location of $^{6}$H g.s.\ as an open question and
provide below some theoretical arguments supporting the g.s.\ prescription based
on the data of our work.


\subsection{Analogies among He and H excitation spectra}


Let us consider the energy level evolution from $^{5}$He (with the assumed
configuration of one neutron particle in the $p_{3/2}$ subshell) to $^{7}$He
(one neutron hole in the $p_{3/2}$ subshell), see Figure \ref{fig:analogy}. The
$3/2^-$ ground state of $^{7}$He becomes more bound, than that in
$^{5}$He. The experimental status of the $1/2^-$  state in $^{7}$He is not well
established (see e.g.\ \cite{Renzi:2016}), however, it seems to have higher
excitation energy than the $1/2^-$ state in $^{5}$He. Moreover, it is highly
likely that there is the $5/2^-$ state in $^{7}$He, built on the $2^+$
excitation of $^{6}$He \cite{Korsheninnikov:1999,Wuosmaa:2008}, which,
evidently, has no counterpart in
$^{5}$He.

\begin{figure}
\begin{center}
\includegraphics[width=0.45\textwidth]{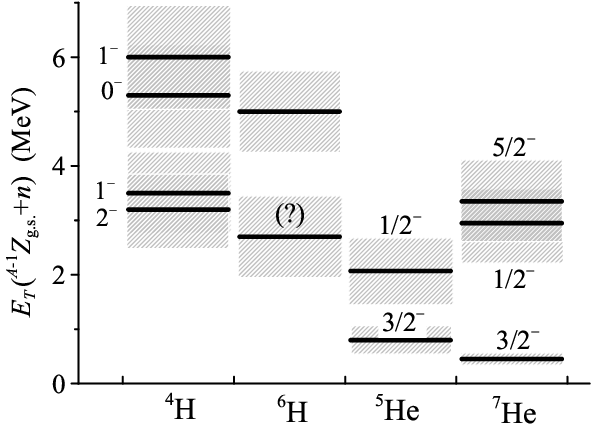}
\end{center}
\caption{Evolution of the level scheme of the $^{5}$He-$^{4}$H pair (one neutron
particle in the $p_{3/2}$ subshell) and the $^{7}$He-$^{6}$H pair (one neutron
hole in the $p_{3/2}$ subshell). The energies are calculated relatively to the
one-neutron separation threshold (the $^{5}$H g.s.\ is assumed to be 1.8 MeV
above the two-neutron separation threshold
\cite{Korsheninnikov:2001,Golovkov:2005}).}
\label{fig:analogy}
\end{figure}

If we consider evolution from $^{5}$He to $^{4}$H, than the $\{3/2^-,1/2^-\}$
spin-orbit doublet is replaced by a quadruplet $\{2^-,1^-,1^-,0^-\}$ of states
 due to split induced by the $^{3}$H spin. If we extend the $^{5}$He-$^{7}$He
analogy of Fig.\ \ref{fig:analogy} to the $^{6}$H states, then
two effects are expected.

\noindent (i) Following the $^{7}$He vs.\ $^{5}$He analogy, we expect
that the $^{6}$H g.s.\ is more bound than the $^{4}$H ground state.
This assumption is true if the 4.5 MeV state really exists in $^{6}$H.

\noindent (ii) In the range $4<E_T< 9$ MeV we expect \emph{six} states
of $^{6}$H. So, it is highly likely that the broad 6.8 MeV structure is actually
a superposition of several overlapping states, which are populated in unknown
proportions and could not be resolved in the inclusive (no correlation)
experiment. It still makes sense to distinguish the 4.5 MeV state, as the lowest
energy resonance, which is allowed by our data, and,
thus, is a candidate to represent the $^{6}$H ground state.


\subsection{Paring energy}


As we have mentioned in Introduction, the $^{6}$H g.s.\ position was suggested
to be at $E_T = 2.6 - 2.9$ MeV in Refs.\
\cite{Aleksandrov:1984,Belozyorov:1986,Caamano:2008}. However, now the g.s.\
energies are known for $^{5}$H ($E_T\approx 1.8$ MeV
\cite{Korsheninnikov:2001,Golovkov:2005}; the $E_T=2.4(3)$ MeV value from
\cite{Wuosmaa:2017} is practically consistent with this value) and $^{7}$H
($E_T\approx 2.2$ MeV \cite{Muzalevskii:2021}; we regard the $E_T\sim 0.3-1$ MeV
value from \cite{Caamano:2008} as much less reliable). Based on these values,
the energy reported in \cite{Aleksandrov:1984,Belozyorov:1986,Caamano:2008} for
the $^{6}$H ground state, would mean the lack of the neutron pairing effect in
the even-neutron nucleus $^{7}$H (experimental paring energy appears to be $\sim
0.7 - 1$ MeV compared with $\sim 3$ MeV expected in analogy with the
$^{7}$He-$^{8}$He pair). Hence, we conclude that the results reported in Refs.\
\cite{Aleksandrov:1984,Belozyorov:1986,Caamano:2008} are not compatible with the
standard pairing assumption. The $^{6}$H ground state suggested in this work at
$E_T = 4.5$ MeV precisely fits the pairing energy systematics.


\subsection{Strong $nn$ correlation observed in pion double charge exchange}


The search for $^{6}$H in the $^{6}$Li($\pi^-$,$\pi^+$) reaction provided no
g.s.\ identification in the $E_T=0-5$ MeV range \cite{Parker:1990}. However, the
authors have pointed in a dedicated paper \cite{Seth:1991} that the peculiar
behavior of the low-energy $^{6}$H missing mass spectrum can be understood as
connected with the presence of strongly correlated $2n$ configuration in the
$^{6}$H continuum considered as $^3$H+$n$+$2n$. We should emphasize that this
observation is actually consistent with the observation of the strong $n$-$n$
correlation in the $^{6}$H decay made in this work, and in $^{5}$H decay in
Ref.\ \cite{Golovkov:2005}.


\section{Conclusions}


The $^{6}$H spectrum was populated in this work in the
$^2\text{H}(^8\text{He},{^4\text{He}})^{6}$H transfer reaction. The broad bump
in the $^{6}$H MM spectrum at $E_T=4-8$ MeV is reliably identified in the data
with the population cross section $d\sigma/d\Omega_{\text{c.m.}}\simeq
190^{+40}_{-80}$
$\mu$b/sr in the $5^{\circ}<\theta_{\text{c.m.}}<16^{\circ}$ angular range. This
is reasonably large cross section, consistent with the resonant population
mechanism. This bump can be interpreted as a broad ($\Gamma>5$ MeV) resonant
state at $E_T=6.8(5)$ MeV. Actually this could be either a single state or a set
of broad overlapping $p$-wave states, as expected from analogy with the known
$^{4}$H spectrum. Observation of such a states(s) is consistent with the data of
Ref.\ \cite{Gurov:2007}, concerning the lowest $^{6}$H state.

We have found \emph{no evidence} of the $\approx 2.6-2.9$ MeV state in $^{6}$H,
which was reported in the pioneering work \cite{Aleksandrov:1984} and has got
support in \cite{Belozyorov:1986,Caamano:2008}. The cross section limit
$d\sigma/d\Omega_{\text{c.m.}} \lesssim 5$ $\mu$b/sr is set for the population
of possible states with $E_T<3.5$ MeV. Also the existence of the $^{6}$H g.s.\
at $\approx 2.6-2.9$ MeV is hardly consistent, due to the pairing energy 
argument,
with the observation of the $^{7}$H g.s.\ at 2.2(5) MeV \cite{Muzalevskii:2021}.
There is no sensible structural argument explaining why the population of the
possible $\approx 2.6-2.9$ MeV ground state could be suppressed in a very
``simplistic'' $^2\text{H}(^8\text{He},{^4\text{He}})^{6}$H transfer reaction
and not observed in our data, while the $^{6}$H spectrum at $E_T \gtrsim 3.5$
MeV is well populated. Therefore, we suggest that position of the $^{6}$H g.s.\
is not yet established, and discussion of this issue should be continued.

The broad bump in the $^{6}$H MM spectrum at $E_T=4-8$ MeV can also be
interpreted as overlap of two relatively narrow states. Such an interpretation
of the experimental spectrum allows us to establish $E_T=4.5(3)$ MeV, as the
\textit{low-energy limit} for the $^{6}$H ground state energy admissible by our
data. According to the energy systematics and the paring energy arguments,
resonance with such an energy is a good candidate for the $^{6}$H ground state.

The low-energy limit of the $^{6}$H g.s.\ position, established as $E_T=4.5(3)$
MeV, confirms that the decay mechanism of the $^{7}$H g.s.\ (located at 2.2 MeV
above the $^{3}$H+$4n$ threshold \cite{Muzalevskii:2021}) is the ``true'' (or
simultaneous) $4n$ emission. Thus, the $^{7}$H is the first confirmed case of
nucleus possessing this exclusive few-body dynamics of decay.

The momentum distribution of the $^{3}$H decay fragments was reconstructed in
the $^{6}$H rest frame. In this work the theoretical studies of the four-body
sequential $^{6}$H$\, \rightarrow \, ^5$H(g.s.)+$n \, \rightarrow \,^3$H+$3n$
decays were performed for the first time. Within the assumption of the $^{6}$H
sequential decay we have found that our data provide evidence that an
extremely strong ``dineutron-type'' correlation is realized in the decay
of the $^{5}$H ground state. More accurate measurements are needed for
more solid conclusions. However, a potentially powerful approach for extracting
information about the nuclear decay dynamics is already illustrated in our work.

It is clear that our work paves a way to more detailed studies of the
$^2\text{H}(^8\text{He},{^4\text{He}})^{6}$H reaction, which would be able to
provide unequivocal results on the excitation spectrum of $^{6}$H.


\acknowledgments


This work was partly supported by the Russian Science Foundation grant No.\
22-12-00054. I.A.M.\ was supported by the Student Grant Foundation of the
Silesian University in Opava, Grant No.\ SGF/2/2020, which was realized within
the EU OPSRE project entitled ``Improving the quality of the internal grant
scheme of the Silesian University in Opava'', No.\
CZ$.02.2.69/0.0/0.0/19\_073/00116951$. The authors are grateful to Prof.\ M.S.\
Golovkov for help in the
work and useful remarks. We acknowledge the interest and support of this
activity from Profs.\ Yu.Ts.\ Oganessian and S.N.\ Dmitriev.


\bibliographystyle{apsrev4-1}
\bibliography{d:/latex/all}


\end{document}